\documentclass[10pt,twocolumn,reqno, aps,prx,superscriptaddress]{revtex4-2}%
\usepackage{eurosym}
\usepackage{amsmath}
\usepackage[pdftex]{graphicx}
\usepackage{float}
\usepackage{color}
\usepackage{rotating}
\usepackage[breaklinks=true,colorlinks,citecolor=blue,linkcolor=blue,urlcolor=blue]%
{hyperref}
\usepackage{graphicx}
\usepackage{natbib}
\usepackage{dcolumn}
\usepackage{bm}
\usepackage{hyperref}
\usepackage[caption=false]{subfig}
\usepackage{epstopdf}
\usepackage{ulem}
\usepackage{amsfonts}
\usepackage{amssymb}%
\providecommand{\U}[1]{\protect\rule{.1in}{.1in}}
\providecommand{\U}[1]{\protect\rule{.1in}{.1in}}
\providecommand{\U}[1]{\protect\rule{.1in}{.1in}}
\providecommand{\U}[1]{\protect\rule{.1in}{.1in}}
\begin{document}

\title{Dynamic control of random telegraph noise in magnetic tunnel junctions}
\author{Mehrdad Elyasi}
\email{mehrdad.elyasi.b2@tohoku.ac.jp}
\affiliation{Advanced Institute for Materials Research, Tohoku University, Sendai 980-8577, Japan}
\affiliation{Center for Science and Innovation in Spintronics, Tohoku University, 2-1-1 Katahira, Sendai 980-8577, Japan.}
\author{Shun Kanai}
\affiliation{Laboratory for Nanoelectronics and Spintronics, Research Institute of Electrical Communication, Tohoku University, 2-1-1 Katahira, Sendai 980-8577, Japan}
\affiliation{Advanced Institute for Materials Research, Tohoku University, Sendai 980-8577, Japan}
\affiliation{Center for Science and Innovation in Spintronics, Tohoku University, 2-1-1 Katahira, Sendai 980-8577, Japan.}
\affiliation{PRESTO, Japan Science and Technology Agency (JST), Kawaguchi 332-0012, Japan. }
\affiliation{Division for the Establishment of Frontier Sciences of Organization for Advanced Studies at Tohoku University, Tohoku University, Sendai 980-8577, Japan.}
\affiliation{National Institutes for Quantum Science and Technology.}
\author{Hideo Ohno}
\affiliation{Center for Science and Innovation in Spintronics, Tohoku University, 2-1-1 Katahira, Sendai 980-8577, Japan.}
\affiliation{Center for Innovative Integrated Electronic Systems, Tohoku University, 468-1 Aramaki Aza Aoba, Sendai 980-8572, Japan.}
\author{Shunsuke Fukami}
\affiliation{Laboratory for Nanoelectronics and Spintronics, Research Institute of Electrical Communication, Tohoku University, 2-1-1 Katahira, Sendai 980-8577, Japan}
\affiliation{Advanced Institute for Materials Research, Tohoku University, Sendai 980-8577, Japan}
\affiliation{Center for Science and Innovation in Spintronics, Tohoku University, 2-1-1 Katahira, Sendai 980-8577, Japan.}
\affiliation{Center for Innovative Integrated Electronic Systems, Tohoku University, 468-1 Aramaki Aza Aoba, Sendai 980-8572, Japan.}
\affiliation{Inamori Research Institute of Science, Shijo, Shimogyo-ku, Kyoto 600-8411, Japan.}
\author{Gerrit E. W. Bauer}
\affiliation{Advanced Institute for Materials Research, Tohoku University, Sendai 980-8577, Japan}
\affiliation{Institute for Materials Research, Tohoku University, Sendai 980-8577, Japan}
\affiliation{Center for Science and Innovation in Spintronics, Tohoku University, 2-1-1 Katahira, Sendai 980-8577, Japan.}

\begin{abstract}
Faster random telegraph noise (RTN) in magnetic tunnel junctions (MTJs) would be beneficial for probabilistic computing applications. However, the interactions between the macrospin and spin waves with finite wave numbers reduce the RTN attempt frequency. We theoretically show that mode-selective heating and cooling by parametric excitation of Kittel mode or propagating spin waves can substantially increase or decrease the RTN frequency, respectively, and propose a nonlinear cooling mechanism that accelerates the switching dynamics. We outline experimental pathways to characterize the nonlinear processes that maximize the operation speed of MTJ-based probabilistic (p-) bits.
\end{abstract}
\maketitle
\date{\today}
\section{\label{Introduction} Introduction}
The random telegraph noise (RTN) of magnetic devices arising from stochastic switching \cite{Neel1949,Brown1963,Brown1979,Braun1993,Braun1994_1,Krause2009,Hayakawa2021,Kanai2021,Safranski2021} has found application in probabilistic computing \cite{Camsari2017,Borders2019,Vodenicarevic2017,Mizrahi2018,Kaiser2022,Singh2024,Si2024}. 
Probabilistic (p-) bits operate by the RTN associated with the switching of the magnetization of the free layer in magnetic tunnel junctions (MTJs) which can be scaled down to cross sections of a few nanometers \cite{Watanabe2018,Jinnai2020,Igarashi2024}. The stochastic switching between parametrically driven bistable precessional magnetization states offers an alternative realization of RTN in magnetic systems \cite{Makiuchi2021,Hioki2021,Elyasi2022}. 
On the other hand magnons, the quanta of spin wave excitations of the magnetic order, are at the brink of entering the quantum regime \cite{Lachance2020,Elyasi2020,Yuan2022,Rameshti2022,Chumak2022,Hioki2026}. The large magnetization fluctuations associated with the magnetization reversal and the small fluctuations of magnons are not independent.

Brown \cite{Brown1963,Brown1979} addressed the distribution function of the macrospin directions around minima of the free energy and computed the switching frequencies using Kramers' escape approximations \cite{Kramers1940}. The average switching time \(\tau_s\)  or frequency $f_{s}=1/\tau_{s}$ can be described by the Néel-Arrhenius Law $\tau_{s}=\tau_{0}e^{E_{B}/(k_{B}T_{\textrm{env}})}$ \cite{Arrhenius1889,Neel1949}, where  $\tau_{0}$ is the attempt time, $E_{B}$ is the barrier energy, $k_{B}$ is the Boltzmann constant, and $T_{\textrm{env}}$ is the global temperature. 

The dynamics of a stochastic switching event depends on the size and shape of the magnetic particle. When the size is much larger than a domain wall width, the complex process of nucleation and motion of magnetic domain walls triggers the reversal. In small enough magnetic particles domain walls do not form, but their order parameters are not infinitely rigid either. We were inspired by experiments \cite{Kanai2023,Kaneko2024} that reported a slow-down of the attempt frequency in samples that do not allow for domain wall nucleation that could not be explained by the macrospin dynamics. We argued that the nonlinear interaction of the macrospin with spin waves acts as a brake on the magnetization RTN  \cite{Elyasi2024}. Since the effect depends on the amplitude of the thermal magnon fluctuations, we hypothesized that tuning their (effective) temperature could help to increase $\tau_{0}$ again.

Here we report that actively manipulating magnon fluctuations by microwaves complements passive methods such as tuning the magnetic field or anisotropy \cite{Hayakawa2021,Endean2014,Talatchian2021,Soumah2024} to engineer the switching dynamics. 
Parametric excitation of the Kittel mode or macrospin  enhances $f_{s}$ [Sec. \ref{Kittel_param}], while exciting spin waves with finite momentum decreases it [Sec. \ref{k_param}]. The in-situ mode-selective cooling mechanism introduced in Sec. \ref{nonlin_cool_model}, also increases $f_{s}$ according to Sec. \ref{nonlin_cool_results}. In Sec. \ref{exp}, we suggest concrete experiments that can test our predictions. 

\section{\label{heating} Parametric excitation and magnetization RTN}
Parametric excitation below a threshold power squeezes and suppresses spin wave fluctuations. On the other hand, above-threshold parametric excitation excites  coherent magnetization precession  that can be in-phase or out-of-phase with the microwave drive with large cone angles \cite{Kinsler1991,Carmichael2,Walls2008,Hioki2021,Elyasi2022}. Analogous to MTJs in which the magnetization RTN occurs between two degenerate states at the poles of the Bloch sphere, the bistability leads to random telegraph noise by the stochastic switching between the two phases \cite{Makiuchi2021,Elyasi2022}. Here, we establish a connection between the fluctuations of these two types of probabilistic bits (p-bits). We focus on the experimentally most relevant thin-film magnetic discs with out-of-plane hard-axis and in-plane easy axis anisotropies.

\begin{figure*}[ptb]
\includegraphics[width=1\textwidth]{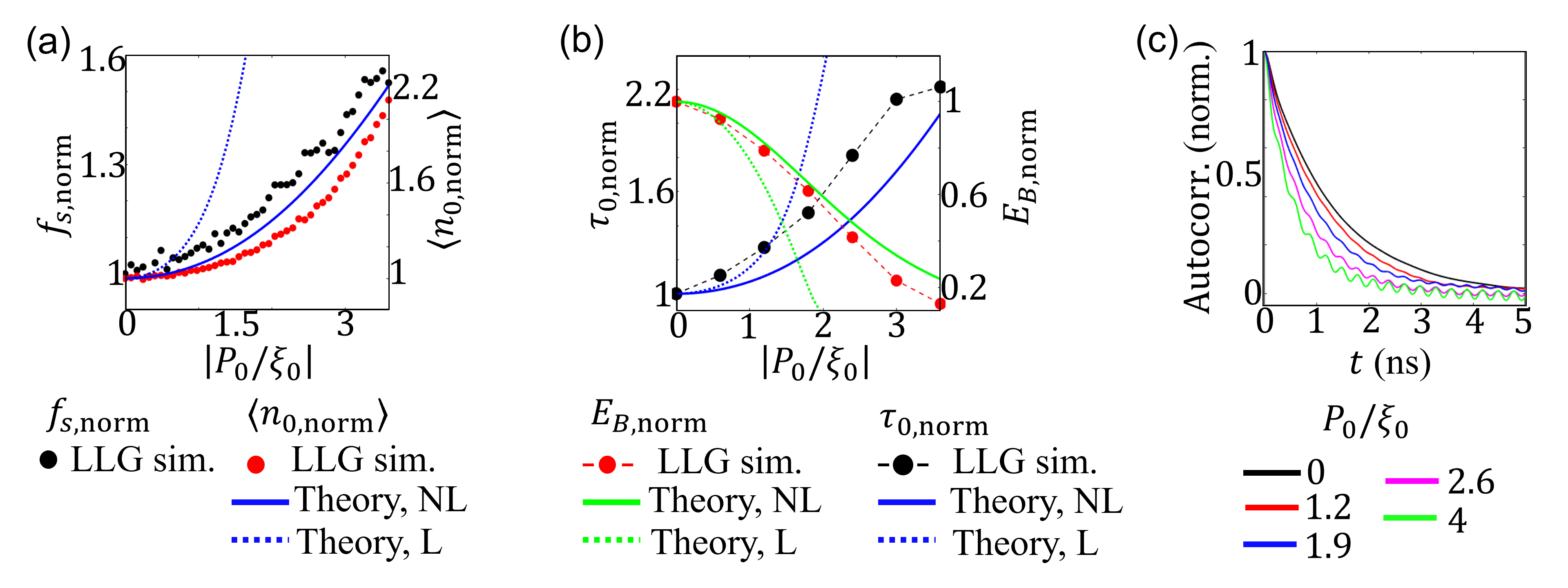}\caption{Parametric excitation and RTN dynamics in the macrospin model. (a) {$f_{s,\textrm{norm}}$ (left axis) and $\langle n_{0,\textrm{norm}}\rangle$ (right axis) are the switching rate and magnon occupation numbers as a function of the microwave drive power $P_{0}$ divided by the damping rate \(\xi_0\), normalized to their equilibrium values at  $P_{0}=0$. (b) Similarly, the switching time $\tau_{0,\textrm{norm}}$ (left axis) and energy barrier $E_{B,\textrm{norm}}$ (right axis). In (a) and (b) `LLG sim.' labels the results obtained numerically by the stochastic LLG equation, while `Theory, L' and `Theory, NL' denote the results of the theoretical model without and with inclusion of the self-Kerr nonlinearity, respectively.} (c) Calculated
autocorrelation normalized to the maximum absolute value, as a function of
time, for different values of $P_{0}$. In (a)-(c), only the macrospin is parametrically excited. {$f_{s}=0.26\,$GHz at $P_{0}=0$. (a)-(c) are the numerical results of stochastic LLG dynamics for a sample with a diameter of $20\,$nm, while the in-plane angle of equilibrium magnetization with easy axis $\theta_{s}=0$.}}
\label{fig1}%
\end{figure*}

\subsection{\label{Kittel_param} Macrospin excitation}
Before addressing nonlinear interactions with spin waves (see Sec. \ref{k_param}), we set the stage by considering the RTN of a rigid macrospin model that allows an analytic treatment with an accuracy that can be assessed by  numerically exact calculations. In a high energy barrier approximation, Brown derived the attempt time \cite{Brown1979}
\begin{align}
\tau_{0}=F(\alpha_G,\lambda_{sd,\theta},\lambda_{sd,\phi})[\sqrt{\lambda_{i,\phi}\lambda_{i,\theta}}]^{-1},\label{eq_param2}
\end{align}
where the curvatures at the saddle points $\lambda_{x,y}=(\partial^2E/\partial y^2)\vert_{\theta_{x}}$, $x\in\{sd,1,2\}$, $y\in\{\theta,\phi\}$. $(\theta,\phi)$ are the azimuthal and polar angles of the magnetic moment relative to the easy axis. `sd' refers to the saddle point of the free energy $E(\theta,\phi)$ of the macrospin, while $i\in\{1,2\}$
selects one of the two equivalent energy minima. The
function $F$ also depends on the Gilbert damping constant $\alpha_{G}$, while the values of
$\lambda_{x,y}$ depend on $M_{s}$ and the anisotropy constants.

MTJs can be parametrically excited either directly by microwave photons or by means of fast actuation of the anisotropy [see Sec. \ref{exp}]. Here we adopt a microwave drive with magnetic field component parallel to the equilibrium magnetization and the easy axis, which efficiently drives the magnetization close to the equilibrium. During the stochastic switching process, the magnetization takes paths over the energy barrier, at which the free energy has a saddle point. At the saddle point, the magnetization is then perpendicular to the microwave magnetic field with frequencies far from the resonance. Consequently, we may disregard the effects of the microwaves on the saddle point dynamics. 

{The non-interacting macrospin Hamiltonian in the absence of the drive reads
\begin{align}
\mathcal{H}_{\mathrm{m}}=\hbar\gamma\mathcal{A}_{0}a_{0}^{\dagger}a_{0}+\hbar\gamma{\mathcal{B}_{0}}(a_{0}^{\dagger 2}+a_{0}^2),\label{eq_param3_1X}
\end{align}
where $\mathcal{A}_{0}=H_{K_{e}}+H_{K_{h}}/2$, $\mathcal{B}_{0}=H_{K_{h}}/4$, $H_{K_{e(h)}}=2K_{e(h)}/M_{s}$, $K_{e(h)}$ is the easy (hard) axis anisotropy, and $a_{0}$ is the  annihilation operator of the uniform excitation. Diagonalization of $\mathcal{H}_{\mathrm{m}}$ leads to $\mathcal{H}_{m'}=\hbar\omega_{0}c_{0}^{\dagger}c_{0}$, where $\omega_{0}=\sqrt{\mathcal{A}_{0}^2-(2\mathcal{B}_{0})^2}$ and $c_{0}=u_{0}a_{0}-v_{0}a_{0}^{\dagger}$ are the frequency and the annihilation operator of the Kittel mode, respectively. $u_{0}=\sqrt{(\mathcal{A}_{0}/\omega_{0}+1)/2}$ and $v_{0}=\sqrt{u_{0}^2-1}$. $u_{0}>1$ indicates the magnon ellipticity that allows parametric excitation. }

{The Fokker–Planck equation governs the thermal distribution function of the magnetization directions \cite{Brown1963,Brown1979}. When disregarding non-linearities, the distribution function of the macrospin model that solves the  Fokker–Planck equation around the free energy minimum reads }
\begin{align}
W_{0}=\mathcal{N}^{\prime}\exp[-2(X_{0}^2+Y_{0}^2)/n_{th,0}], \label{eq_param2_1}
\end{align}
where $\mathcal{N}^{\prime}$ is a normalization constant, $X_{\vec{k}}=(c_{\vec{k}}+c^{\dagger}_{\vec{k}})/2$, $Y_{\vec{k}}=-i(c_{\vec{k}}-c^{\dagger}_{\vec{k}})/2$,  $c_{\vec{k}}$ is the annihilation operator of the magnon mode with wave vector $\vec{k}$, and $n_{th,\vec{k}}$ is the thermal occupation at the bath temperature.

A microwave magnetic field polarized parallel to the easy axis then
contributes the Hamiltonian $\mathcal{H}_{p}=-P_{0}\sin{(\omega_{p}t)}a_{0}^{\dagger}a_{0}$,  where $P_{0}$ is a real number in units of frequency that is proportional to the applied microwave power. In the rotating frame of the Kittel mode frequency and an excitation frequency fixed to $\omega_{p}=2\omega_{0}$, the total Hamiltonian of the Kittel mode becomes $\mathcal{H}_{m'}^{T}=(-iP_{p,0}c_{0}c_{0}+\mathrm{H.c.})/2$, where $P_{p,0}=P_{0}u_{0}v_{0}$ and the distribution function in the presence of the drive and dissipation becomes \cite{Kinsler1991,Walls2008}
\begin{align}
&W_{0,p}=\mathcal{N}\exp \left\{ \frac{2}{n_{th,0}}\left[-(1+\epsilon_{0})X_{0}^2-(1-\epsilon_{0})Y_{0}^2\right]\right\},\label{eq_param3}
\end{align}
where $\mathcal{N}$ is a normalization constant, $\epsilon_{0}=P_{p,0}/\xi_{0}$, and $\xi_{0}=(u_{0}^2+v_{0}^2)\alpha_{G}\omega_{0}$ is the Kittel mode dissipation rate. 

{Transforming back to  the laboratory frame: }
\begin{align}
&W_{m,0,p}=\mathcal{N}_{m}\exp\{-\Phi_{p}\}\nonumber\\
&\Phi_{p}=\frac{(u_{0}-v_{0})^2x_{0}^2}{n_{th,0}}[\mathcal{R}\cos^2(\omega_0 t)+\mathcal{R}'\sin^2(\omega_0 t)]\nonumber\\
&+\frac{(u_{0}+v_{0})^2y_{0}^2}{n_{th,0}}(\mathcal{R}\sin^2(\omega_0 t)+\mathcal{R}'\cos^2(\omega_0 t))\nonumber\\
&-\frac{2}{n_{th,0}}[\sin(\omega_0 t)\cos(\omega_0 t)(\mathcal{R}-\mathcal{R}')x_{0}y_{0}]. \label{eq_param2_new1}
\end{align}
{where $\mathcal{R}=1+\epsilon_{0}$, $\mathcal{R}'=1-\epsilon_{0}$, $x_{0}=(a_{0}+a_{0}^{\dagger})/2$, and $y_{0}=-i(a_{0}-a_{0}^{\dagger})/2$. 
}

{According to Kramers' escape theory, the escape rate or switching frequency is proportional to the probability of finding particles near the barrier. The prefactor $\tau_{0}$ is therefore inversely proportional to $\mathcal{N}_{m}$ \cite{Brown1979,Braun1994_1,Elyasi2024}.} {At equilibrium}, $\tau_{0,0}\propto(\lambda_{1(2),\theta}\lambda_{1(2),\phi})^{-1/2}${, see Eq. (\ref{eq_param2}), while}
{in the presence of the drive}
\begin{align}
&\frac{\tau_{0,P_{p,0}}}{\tau_{0,0}}=\frac{\mathcal{N}_{m}\vert_{0}}{\mathcal{N}_{m}\vert_{P_{p,0}}}=\nonumber\\
&\frac{\int_{-\inf}^{\inf}\exp[-\Phi_{P}\vert_{P_{p,0}}] dx_{0}dy_{0}}{\int_{-\inf}^{\inf}\exp[-\Phi_{P}\vert_{0}]dx_{0}dy_{0}}=\frac{1}{\sqrt{1-\epsilon_0^2}},\label{eq_param2_new2}
\end{align}
{where $\vert_{P_{p,0}}$ denotes evaluation at $P_{p,0}$. $\tau_{0,P_{p,0}}$ therefore appears to increase under parametric excitation.  } 

{However, this is not the whole story since the microwaves also affect the effective potential barrier by shifting the minima relative to the saddle point \cite{Devoret1984}. The Kramers' escape rate follows from the effective time-independent probability of finding a particle at the barrier, }
\begin{align}
W_{m,0,p,\textrm{eff}}=\mathcal{N}_{m}\exp\left[{\frac{-x_{B}^2}{2\langle\langle x_{0}^2\rangle\rangle_{t}}}\right],\label{eq_param2_new3}
\end{align}
{where the symbols $\langle \cdot \rangle$ and $\langle \cdot \rangle_{t}$ denote statistical and time averages, respectively. Since the probability current is along the line connecting the minima to the saddle node, i.e., phase space variable $x_{0}$, we used $y_{B}=0$. The switching time in the presence of drive }
\begin{align}
\tau_{s,P_{p,0}}=\tau_{0,P_{p,0}}\exp \left[\frac{E_{B,P}}{k_{B}T}\right],\label{eq_param3_1}
\end{align}
{where $E_{B,P}=E_{B,0}-E_{P}$ and $E_{B,0}$ is the barrier energy without parametric excitation. From Eqs. (\ref{eq_param2_new3}) and (\ref{eq_param3_1})},
\begin{align}
\frac{\tau_{s,P_{p,0}}}{\tau_{s,0}}=\frac{W_{m,0,p,\textrm{eff}}\vert_{0}}{W_{m,0,p,\textrm{eff}}\vert_{P_{p,0}}}=\frac{\tau_{0,P_{p,0}}}{\tau_{0,0}}\exp\left[{\frac{-E_{P}}{k_{B}T}}\right]
\end{align}
where
\begin{align}
E_{P}=E_{B,0}^{\prime}\left(1-\frac{\langle\langle x_{0}^2\rangle\rangle_{t}\vert_{0}}{\langle\langle x_{0}^2\rangle\rangle_{t}\vert_{P_{p,0}}}\right),\label{eq_param2_new4}
\end{align}
{$E_{B,0}\approx E_{B,0}^{\prime}  =k_BT{\frac{x_{B}^2}{2\langle\langle x_{0}^2\rangle\rangle_{t}}}$ is a parabolic estimate.}
{Averages over the distribution $W_{m,0,p}$ lead to}
\begin{align}
&\langle\langle x_{0}^2\rangle\rangle_{t}= \frac{n_{th,0}}{2(u_0-v_0)^2(1-\epsilon_0^2)} \label{eq11}
\end{align}
\begin{align}
&\langle\langle n_{0}\rangle\rangle_{t}=\frac{n_{th,0}}{1-\epsilon_0^2}, \label{eq12}
\end{align}
{which with Eq. (\ref{eq_param2_new4}) gives}
\begin{align}
E_{P}=E_{B,0}\epsilon_{0}^2.\label{eq_param2_new5}
\end{align}
{ $\tau_{s}$ decreases when switching on the drive when $\tau_{0,P_{p,0}}/\tau_{0}<\exp[E_{P}/(k_{B}T)]$ or $T>E_{p}-\ln{(\tau_{0,P_{p,0}}/\tau_{0})}$, i.e., when the exponential term dominates. A divergence at the threshold $P_{p,0}=\xi_{0}$ is unphysical because the analytical expressions for $\tau_{0,P_{p,0}}$ and $E_{P}$ only hold for small $\epsilon_{0}$.  Nevertheless, the arguments remain qualitatively valid because nonlinearities regulate $\Phi_{P}$, as confirmed by simulations as follows.}

We assess the validity of Kramers' escape model of the switching time by solving the stochastic LLG equation for the macrospin as a function of $P_{0}/\xi_{0}$ under parametric microwave excitation. We start from an equilibrium magnetization governed by in-plane (IP) easy axis and out-of-plane (OOP) hard-axis anisotropies with $H_{K_{h}}=0.5\,$T and $H_{K_{e}}=-7.5\,$mT for a  saturation magnetization $M_{s}=1.5\times 10^6\,$A/m, which is typical for CoFeB MTJs \cite{Hayakawa2021,Kanai2023}. We adopt a Gilbert damping constant $\alpha_{G}=0.05$ and a film thickness of $2\,$nm, unless stated otherwise. \(K_e\) accounts for the elliptical shape of the experimental MTJ discs that we then take to be circular discs with radius $10\,$nm. Our method can be easily extended to magnets with equilibrium magnetizations forced OOP by a net perpendicular crystal anisotropy \cite{Hayakawa2021}. 

Figure \ref{fig1}(a) illustrates the increase of $f_{s,\textrm{norm}}$ and $\langle n_{0,\textrm{norm}}\rangle$ with $\vert P_{0}\vert$, where the subscript `norm' refers to a normalization relative to their equilibrium values. According to Figure \ref{fig1}(b), $\tau_{0,\textrm{norm}}$ increases while $E_{B,\textrm{norm}}$ decreases with $\vert P_{0}\vert$ as anticipated. The switching rate increases because the increase in the attempt frequency is overcompensated by the lowering of the activation potential in the exponential term.

{We calculate switching rates by counting the number of switching events and divide it by the total simulation time. In order to assess the presence of coherent oscillations of the magnetization between the two directions, we also computed the time autocorrelation of the macrospin
magnetization along the easy axis $\hat{x}$, i.e., $\mathcal{R}(t)=\langle m_{x}(t')m_{x}(t'+t)\rangle_{t'}$, as a function of $P_{0}$. Figure \ref{fig1}(c) proves that $P_{0}$ does not excite a ``ringing" magnetization. The enhanced RTN frequency $f_{s}$ is consistent with the increasing decay of
the autocorrelation. The oscillatory features at relatively large $P_{0}$
represent the dynamics within the energy wells and is not associated with magnetization switches.}

{In order to understand the physics and parameter dependence of the ``numerically exact" results aided by our analytical theory, we must first ascertain that the approximations in the latter are valid. The dashed lines in Figs. \ref{fig1}(a) and (b) represent Eqs. (\ref{eq_param2_new2}) and (\ref{eq12}) where $u_{0}=1.6$ and $v_{0}=1.24$ for the anisotropy parameters introduced above. While representing the trends correctly, the disagreement with the numerical results increases drastically with $P_{0}$ because of the unphysical divergence at the threshold power. We can improve the theory by including the leading-order nonlinearity, i.e., the self-Kerr interaction, $\mathcal{H}_{MS,NL}=\mathcal{D}^{(\mathrm{Kerr})}_{0}n_{0}^{2}$  treating \(\mathcal{D}^{(\mathrm{Kerr})}_0\) as a parameter that can be adjusted to model higher order non-linearities.  The  master equation of the driven motion of the density matrix $\dot{\rho}=\mathcal{L}\rho$, where $\mathcal{L}=-i[\mathcal{H}^{T}_{m'}+\mathcal{H}_{MS,NL},\rho]+L_{0}$, $L_{0}$ is the linear dissipation operator, must be computed numerically \cite{Elyasi2022, Elyasi2024}.} {The resulting density matrix in the steady-state, $\rho_{ss}$, can be used to compute observables $\langle \hat{O}\rangle=\mathrm{tr}[\rho_{ss}\hat{O}]$, i.e., $\langle n_{0}\rangle$, $\langle X_{0}^2\rangle$, $\langle Y_{0}^2\rangle$. Following Eqs. (\ref{eq_param3})–(\ref{eq_param2_new3}) :}
\begin{align}
&\frac{\tau_{0,P_{p,0}}}{\tau_{0,0}}=\left[ \frac{\langle X_{0}^2\rangle\vert_{P_{p,0}}\langle Y_{0}^2\rangle\vert_{P_{p,0}}}{\langle X_{0}^2\rangle\vert_{0}\langle Y_{0}^2\rangle\vert_{0}}\right]^{1/2},\label{eq_param2_new6}\\
&E_{P}=E_{B,0}\left[1-\frac{\langle X_{0}^2\rangle\vert_{0}\langle Y_{0}^2\rangle\vert_{0}}{\langle X_{0}^2\rangle\vert_{P_{p,0}}\langle Y_{0}^2\rangle\vert_{P_{p,0}}}\right]
\end{align}
{where $\vert_{P_{p,0}}$ and $\vert_{0}$ indicate expectation values evaluated with and without the microwave drive, respectively.}

 {For the computed $E_{P}$ and $\tau_{0,P_{p,0}}/\tau_{0,0}$ in Figs. \ref{fig1}(a) and (b) we fitted the self-Kerr coefficient $\mathcal{D}^{(\mathrm{Kerr})}_{0}$ to $\langle n_{0,\mathrm{norm}}\rangle$ of the micromagnetic calculations [red dots in Fig. \ref{fig1}(a)]. $\vert\mathcal{D}^{(\mathrm{Kerr})}_{0}\vert=2.4\times 10^4 \,\mathrm{Hz}$, is much smaller than the directly calculated $6.7\times 10^4\,\mathrm{Hz}$ \cite{Krivosik2010}, which we  attribute to higher-order nonlinearities in the Holstein–Primakoff expansion near the saddle point of the static free energy. The  excellent agreement between the quasi-analytic model and the LLG simulations implies that we caught the essential physics of the microwave driven RTN in small magnetic particles, which also opens strategies to increase $f_{s}$ as presented in the following sections.}

\begin{figure}[ptb]
\includegraphics[width=0.5\textwidth]{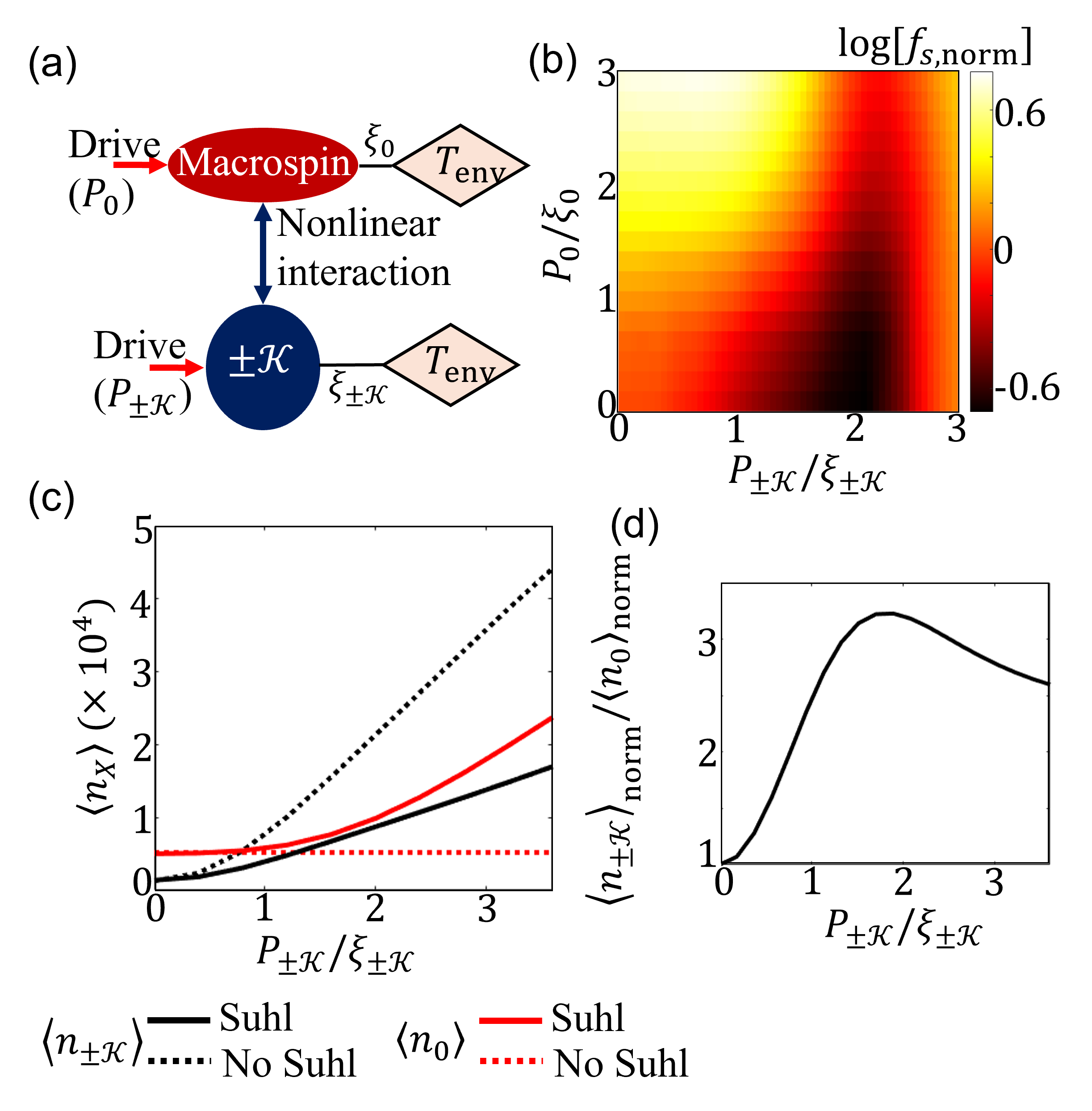}
\caption{RTN dynamics and parametric excitation of the Kittel mode and interacting spin waves. (a) Schematics of the model a  macrospin and magnon pairs with opposite momenta $\pm\mathcal{K}$ driven by microwaves and in contact with two thermal reservoirs. (b) Calculated $\log f_{s,norm}$ as a function of $P_{0}$ and $P_{\pm\mathcal{K}}$ . `norm' in the subscripts refers to the normalization to the equilibrium values ($P_{0}=P_{\pm\mathcal{K}}=0$ at which $f_{s,0}=0.12\,$GHz) to be compared with $f_{s}=0.6\,$GHz for a pure macrospin model. In (b), the sample radius is $20\,$nm with in-plane angle of the equilibrium magnetization with easy axis $\theta_{s}=\pi/4$. {(c) Steady-state $\langle n_{\pm\mathcal{K}}\rangle$ and $\langle n_{0}\rangle$ of the interacting magnon model. Full lines (dashed lines) are in the presence (absence) of the Suhl nonlinear interaction. (d) $\langle n_{\pm\mathcal{K}}\rangle_{\textrm{norm}}/\langle n_{0}\rangle_{\textrm{norm}}$, where $\langle n_{X}\rangle_{\textrm{norm}}$ is normalized by the values at $P_{\pm\mathcal{K}}=0$.}}
\label{fig2}%
\end{figure}

\subsection{\label{k_param} $\pm\vec{k}\neq0$ excitation}
We argued previously \cite{Elyasi2024} that non-uniform spin waves suppress the RTN frequency by correcting the damping of the correlation $\langle c^{\dagger}_{0}c^{\dagger}_{0}\rangle$ (and its hermitian conjugate) of the Kittel mode:
\begin{align}
\xi_{corr}=\frac{2\xi_{\pm\mathcal{K}}\mathcal{D}_{Suhl}^2(2\langle n_{0}\rangle+1)(2\langle n_{\pm\mathcal{K}}\rangle+1)}{\Delta_{\pm\mathcal{K}}^2+\xi_{\pm\mathcal{K}}^2},\label{eq_param3_2}
\end{align}
where $\xi_{\vec{k}}$ is the dissipation rate of a magnon with frequency $\omega_{\vec{k}}$, $n_{\vec{k}}=c^{\dagger}_{\vec{k}}c_{\vec{k}}$ is the number operator, $\mathcal{D}_{Suhl}$ is the strength of the four-magnon interaction $c_{0}c_{0}c_{\mathcal{K}}^{\dagger}c_{-\mathcal{K}}^{\dagger}+\mathrm{H.c.}$ \cite{Suhl1957}, and $\Delta_{\pm\mathcal{K}}=\omega_{\pm\mathcal{K}}-\omega_{0}$. Here, all $\pm\vec{k}\neq 0$ modes have been lumped into a single effective one \cite{Lvov1994,Rezende2020,Lee2023,Makiuchi2024,Elyasi2024} that we refer to as $\pm\mathcal{K}$ pair. Increasing $\langle n_{\pm\mathcal{K}}\rangle$, e.g., by parametric excitation, increases $\xi_{corr}$. A larger $\xi_{corr}$ suppresses the probability current toward the saddle node by the emission of spin waves. At equilibrium, nonlinear interactions do not modify the magnon number and barrier height significantly. We therefore associate the suppression of the switching rate with an increased attempt time $\tau_{0}$, as confirmed by the simulations \cite{Elyasi2024}.  

When there are two thermodynamic baths at temperatures $T_{0}$ and $T_{\pm\mathcal{K}}$,  interacting with the macrospin and $\pm\mathcal{K}$ spin waves, respectively, $f_{s}$ decreases with increasing $T_{\pm\mathcal{K}}$ and constant $T_{0}$ \cite{Elyasi2024}. Here, we demonstrate the control for either of the two temperatures or magnon numbers $\langle n_{\pm\mathcal{K}}\rangle$ and $\langle n_{0} \rangle$ by parametric microwave excitation.

The Hamiltonian of the system with a microwave excitation that parametrically  drives \(\pm\mathcal{K}\)  magnons in the rotating frame of $\omega_{\pm\mathcal{K}}$ reads  \cite{Elyasi2024},
\begin{align}
&\mathcal{H}_{eff,p}=-\Delta_{\pm\mathcal{K}}c_{0}^{\dagger}c_{0}+P_{\pm\mathcal{K}}(c_{\mathcal{K}}c_{-\mathcal{K}}+H.c.)+\nonumber\\
&\mathcal{D}^{(\mathrm{Suhl})}(c_{0}c_{0}c_{\mathcal{K}}^{\dagger}c_{\mathcal{K}}^{\dagger}+\mathrm{H.c.})+\sum_{\vec{k}\in\{0,\pm\mathcal{K}\}}\mathcal{D}^{(\mathrm{Kerr})}_{\vec{k}}c_{\vec{k}}^{\dagger}c_{\vec{k}}c_{\vec{k}}^{\dagger}c_{\vec{k}} ,\label{eq_param4}
\end{align}
where $P_{\pm\mathcal{K}}$ is the drive amplitude at $2\omega_{\pm\mathcal{K}}$, and $\mathcal{D}^{(\mathrm{Suhl})}$ as well as $\mathcal{D}^{(\mathrm{Kerr})}_{\vec{k}}$ are functions of the saturation magnetization $M_{s}$. anisotropies, exchange length, sample thickness, magnetization direction, and wave vector \cite{Krivosik2010}. 

In the absence of an interaction between the Kittel mode and the $\pm\mathcal{K}$ pair, viz. $\mathcal{D}^{(\mathrm{Suhl})}=0$ the equations of motion (EOM) for correlation functions such as $\langle c_{\mathcal{K}}c_{-\mathcal{K}}\rangle$ lead to the (formally exact) transcendental equation for $\langle n_{\pm\mathcal{K}}\rangle$ 
\begin{align}
8{\mathcal{D}^{(\mathrm{Kerr})}_{\pm\mathcal{K}}}^2\langle n_{\pm\mathcal{K}}\rangle^3-8{\mathcal{D}^{(\mathrm{Kerr})}_{\pm\mathcal{K}}}^2n_{th,\pm\mathcal{K}}\langle n_{\pm\mathcal{K}}\rangle^2+\nonumber\\
(2\xi_{\pm\mathcal{K}}^2-2P_{\pm\mathcal{K}}^2)\langle n_{\pm\mathcal{K}}\rangle-2\xi_{\pm\mathcal{K}}^2n_{th,\pm\mathcal{K}}-P_{\pm\mathcal{K}}^2=0.\label{eq_param5}
\end{align}
When $\langle n_{\pm\mathcal{K}}\rangle\gg n_{th,\pm\mathcal{K}}$ and $\langle n_{\pm\mathcal{K}}\rangle\gg1$,
\begin{align}
\langle n_{\pm\mathcal{K}}\rangle \approx \frac{1}{2\vert\mathcal{D}_{\pm\mathcal{K}}^{(\mathrm{Kerr})}\vert}\sqrt{P_{\pm\mathcal{K}}^2-\xi_{\pm\mathcal{K}}^2}.\label{eq_param6}
\end{align}
The predicted continuous increase of $\langle n_{\pm\mathcal{K}}\rangle$ with $\vert P_{\pm\mathcal{K}}\vert$ holds only in the absence of other magnon interactions that kick in at higher excitation powers. Switching on the Suhl interaction $\mathcal{D}^{(\mathrm{Suhl})}$ then drives the Kittel mode via $P_{0,eff}c_{0}c_{0}+$H.c., where
\begin{align}
P_{0,eff}&=\mathcal{D}^{(\mathrm{Suhl})} \langle c_{\mathcal{K}}^{\dagger}c_{-\mathcal{K}}^{\dagger}\rangle\\
&=\mathcal{D}^{(\mathrm{Suhl})}\frac{2iP_{\pm\mathcal{K}}\langle n_{\pm\mathcal{K}}\rangle}{2\xi_{\pm\mathcal{K}}-4i\mathcal{D}^{(\mathrm{Kerr})}_{\pm\mathcal{K}}\langle n_{\pm\mathcal{K}}\rangle},
\label{eq_param7}
\end{align}
Moreover, when $\langle n_{0}\rangle\gg n_{th,0}$, $\langle n_{0}\rangle\gg1$, and $\sqrt{4\vert P_{0,eff}\vert^2-\xi_{0}^2}>\Delta_{\pm\mathcal{K}}$ 
a third-order polynomial equation similar to Eq. (\ref{eq_param5}) leads to
\begin{align}
\langle
n_{0}\rangle=\frac{\sqrt{4\vert P_{0,eff}\vert^2-\xi_{0}^2}-\Delta_{\pm\mathcal{K}}}{2\vert\mathcal{D}_{0}^{(\mathrm{Kerr})}\vert}.
\label{eq_param8}
\end{align}

The driven Kittel mode, vice versa, adds an effective parametric drive on  the $\pm\mathcal{K}$ pair through the Suhl interaction, modifying the total parametric excitation of the magnon pair to  $P_{\pm\mathcal{K}}^{\prime}c_{\mathcal{K}}c_{-\mathcal{K}}+\mathrm{H.c.}$, where 
\begin{align}
P_{\pm\mathcal{K}}^{\prime}=P_{\pm\mathcal{K}}+\mathcal{D}^{(\mathrm{Suhl})}\langle c_{0}^{\dagger}c_{0}^{\dagger}\rangle,\label{eq_param9_1}
\end{align}
\begin{align}
\langle c_{0}^{\dagger}c_{0}^{\dagger}\rangle=\frac{-2iP_{0,eff}^{\ast}\langle n_{0}\rangle}{\xi_{0}-i(\Delta_{\pm\mathcal{K}}-2\mathcal{D}^{(\mathrm{Kerr})}_{0}\langle n_{0}\rangle)}.
\label{eq_param9}
\end{align}
Since here $\Delta_{\pm\mathcal{K}}-2\mathcal{D}^{(\mathrm{Kerr})}_{0}\langle n_{0}\rangle\gg \xi_{0}$, 
$\mathrm{arg}[\langle c_{0}^{\dagger}c_{0}^{\dagger}\rangle]\approx \mathrm{arg}[P_{0,eff}^{\ast}]$.
When increasing $\langle n_{\mathcal{\pm\mathcal{K}}}\rangle$ to $\vert2\xi_{\pm\mathcal{K}}\vert<\vert4\mathcal{D}^{(\mathrm{Kerr})}_{\pm\mathcal{K}}\langle n_{\pm\mathcal{K}}\rangle\vert$, Eq. (\ref{eq_param7}) leads to
$\mathrm{arg}[P_{0,eff}^{\ast}]=\mathrm{arg}[P_{\pm\mathcal{K}}]+\pi$
and Eq. (\ref{eq_param9_1}) is equivalent to $P^{\prime}_{\pm\mathcal{K}}\approx P_{\pm\mathcal{K}}-\vert\mathcal{D}^{(\mathrm{Suhl})}\langle c_{0}^{\dagger}c_{0}^{\dagger}\rangle \vert$. The Suhl interaction therefore effectively suppresses the growth of $\langle n_{\pm\mathcal{K}}\rangle$. 

Summarizing, we predict that by injecting magnons $\langle n_{\pm\mathcal{K}}\rangle$ by increasing $\vert P_{\pm\mathcal{K}}\vert$, $\xi_{corr}$ increases and reduces the switching rate. Above a critical power, the increasing population of the Kittel mode limits the increase of $\langle n_{\pm\mathcal{K}}\rangle$. Moreover, as discussed in Sec. \ref{Kittel_param}, the effective parametric excitation of the Kittel mode also increases $f_{s}$. This competition should lead to a non-monotonic dependence of $f_{s}$ on $\vert P_{\pm\mathcal{K}}\vert$. However, with increasing excitation power, many magnon modes are excited either directly or through nonlinear interactions and our 3-magnon mode model breaks down, as signaled by the micromagnetic calculations presented below.

We numerically assess the above predictions by computing the dynamics of a model in which the up and down macrospin states interact with a $\pm\mathcal{K}$ magnon pair \cite{Elyasi2024} under parametric drives $P_{0}$ and $P_{\pm\mathcal{K}}$ as sketched in Figure \ref{fig2}(a). An appropriate in-plane magnetic field tilts the IP angle of the equilibrium magnetization from the easy axis to $\theta_{s}=\pi/4$ . The disc radius $r=20\,$nm is experimentally feasible and small enough to suppress the nucleation of domain walls. The finite $\theta_{s}$ facilitates switching and thereby accelerates numerical calculations. From all magnons that are represented by the $\pm\mathcal{K}$ pair, those with frequency closest to the Kittel mode dominate, i.e., with one radial node in the plane. The model parameters are then $\omega_{\pm\mathcal{K}}=4.7\,$GHz and Kittel mode $\omega_0=1.2 \,$GHz. The non-linearity parameters $\mathcal{D}^{(\mathrm{Suhl})}=8.1\times10^4\,$Hz and $\mathcal{D}^{(\mathrm{Kerr})}_{\pm\mathcal{K}}=1.6\times10^4\,$Hz follow from the formulas  in Ref. \cite{Krivosik2010} valid in the thin film limit. Figure \ref{fig2}(b) is a plot of the resulting $f_{s,\textrm{norm}}$ on a logarithmic scale. When $P_{\pm\mathcal{K}}$ is fixed, $f_{s}$ increases with $P_{0}$ as found for the macrospin model in Sec. \ref{Kittel_param}. On the other hand, $f_{s}$ depends non-monotonically on $P_{\pm\mathcal{K}}$ for constant $P_{0}$, initially decreasing with $P_{\pm\mathcal{K}}$ followed by an upturn, supporting the qualitative analysis above. {The steady-state occupations of the magnon model (Kittel mode plus a magnon pair) with Hamiltonian $\mathcal{H}_{eff,p}$ [see Eq. (\ref{eq_param4})], presented in Fig. \ref{fig2}(c) illustrates that  the growth rate of $\langle n_{\pm\mathcal{K}}\rangle$ with $P_{\pm\mathcal{K}}$ is suppressed by the Suhl interaction. Figure \ref{fig2}(d) shows that the relative increase of the Kittel mode number surpasses that of the magnon pair at a certain $P_{\pm\mathcal{K}}$ that  coincides with the minimum of $\log[f_{s,\textrm{norm}}]$ in Fig. \ref{fig2}(b).}.

\begin{figure}[ptb]
\includegraphics[width=0.5\textwidth]{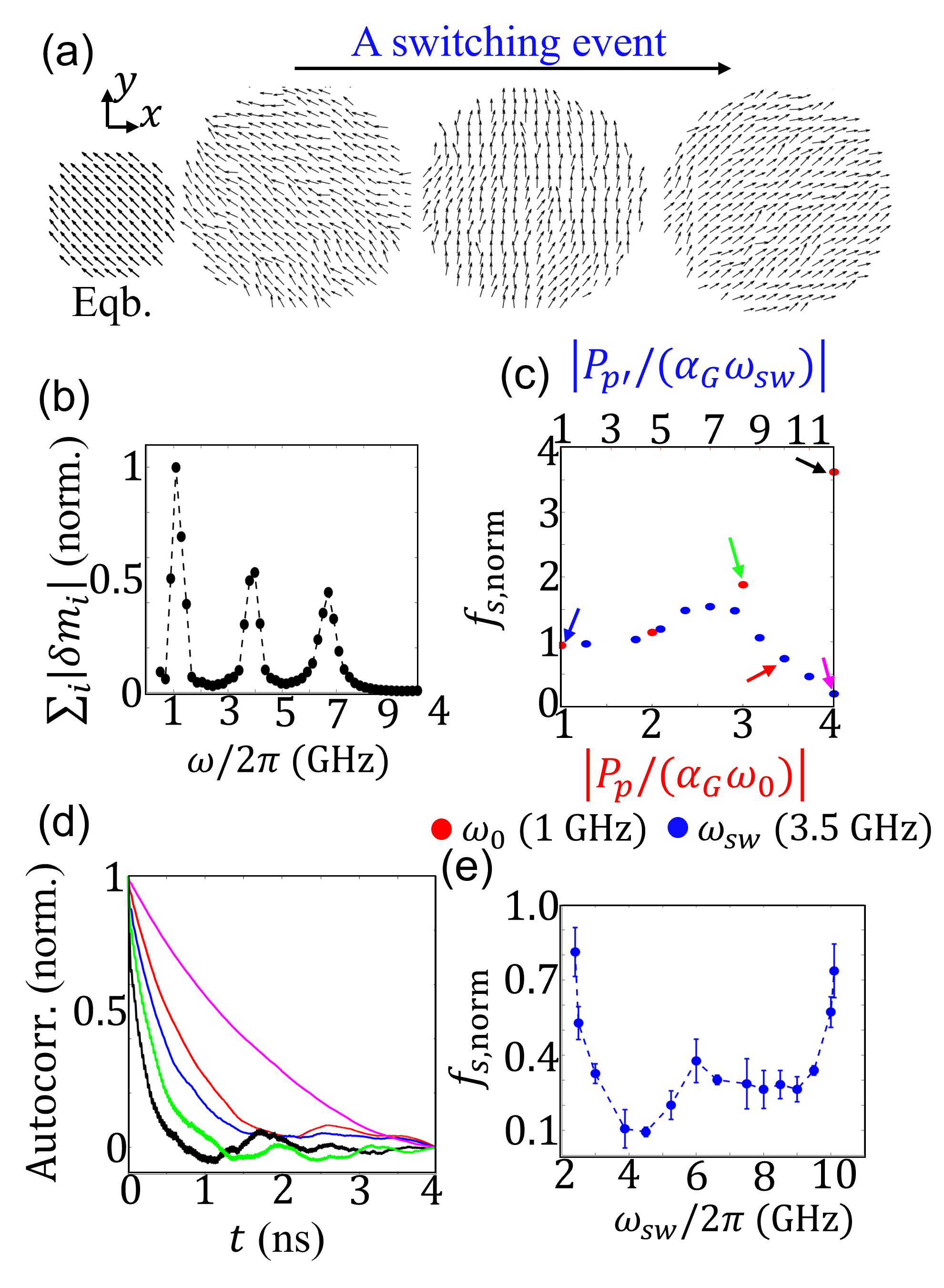}
\caption{{(a)
Representative snapshots of the magnetization texture during a switching
event in the micromagnetic simulations. 
The left-most picture (Eqb.) is the magnetization at zero temperature. {(b) Integrated spin wave amplitude excited by  microwaves with frequency $\omega$ and linearly polarized along the easy axis $\hat{z}$ and linearly varying amplitude along $\hat{x}$. The spectrum is normalized to the maximum amplitude.} (c) $f_{s,\textrm{norm}}$ calculated under a uniform
microwave polarized along $\hat{z}$. The red dots correspond to 
parametric Kittel mode frequency with $P_{p}\neq0$ and
$P_{p^{\prime}}=0$. The blue dots are the results under parametric pumping of the \(\pm\mathcal{K}\) spin waves with $P_{p^{\prime}}\neq0$ while
$P_{p}=0$. Here \(P_p\) (\(P_{p'}\)) is proportional to the power of the microwave drive at \(2\omega_0\) (\(2\omega_{\pm\mathcal{K}}\)). `norm' in the subscript refers to the normalization to
$f_{s,exch-dip}$ when $P_{p}=P_{p^{\prime}}=0$. The colored arrows refer to panel (d). (d) Time-autocorrelation function normalized to unity at $t=0$ for various $P_{p}$ and
$P_{p^{\prime}}$. The colors refer to the switching rates in (c). {(e) $f_{s,\textrm{norm}}$ under parametric excitation of twice the spin wave frequency $\omega_{sw}$ at a fixed $P_{p'}/(\alpha_{G}\omega_{sw})=12$ with statistical error bars.}}}%
\label{fig3}%
\end{figure}

{Our model \cite{Elyasi2024} assumes that the
magnetization switches as a single domain, i.e., without the nucleation of domain walls, while non-uniformity enters only through spin wave oscillations on top of the single-domain dynamics.
Single-domain reversal dominates for tunnel junctions with radii not much larger than the domain wall width  \(r \lesssim \)  $\sqrt
{A_{ex}/|K_{e}|}\sim30\,$nm for exchange stiffness of CoFeB $A_{ex}=1.3\times10^{-11}\,$J/m.  Our model is therefore appropriate to describe the experiments reported by Ref. \cite{Kanai2023}, but not Ref. \cite{Soumah2024}.}

{Numerical calculations of the magnetization dynamics test the model approximations while the model helps to understand numerical experiments.
Here we use a micromagnetic code that solves the stochastic LLG\ equation for a $20\,$nm ($2\,$nm) radius (thickness) disk on a two-dimensional square lattice of magnetic elements with nearest-neighbor exchange and long-range dipolar interactions. 
The latter contributes an out-of-plane hard-axis shape anisotropy
$H_{K_{h,s}}\approx-\mu_{0}M_{s}$. For a better comparison with theory above, we modify the parameter $H_{K_{h}}$ in the micromagnetics such that
$H_{K_{h}}+H_{K_{h,s}}$ equals the anisotropy field used in the model. As above, rather than computing the micromagnetics of elliptic disks we introduce an easy axis anisotropy parameter $H_{K_{e}}$ along $\hat{x}$ in circular
ones [see Fig. \ref{fig3}(a)]. We extract the RTN frequencies from the micromagnetic
dynamics, by averaging 32 runs over $0.8\,\mathrm{\mu}$s and $10^{-14}$s time step. {After confirming that the numbers do not change significantly for a finer in-plane mesh of $2\,\text{nm}\times 2\,\text{nm}$, we computed most results for $3.5\,\text{nm}\times 3.5\,\text{nm}$ unit cells.}
Figure \ref{fig3}(a) shows snapshots of the spatial evolution during a single switching event {for a mesh size of $2\text{nm}\times2\text{nm}$}, confirming a nearly single-domain process with superimposed small spatial non-uniformity, i.e., spin waves. The calculated micromagnetic RTN frequency $f_{s,exch-dip}=0.05\,$GHz is an order of
magnitude smaller than the macrospin RTN frequency $f_{s,MS}=0.6\,$GHz. The simulations therefore support our previous conclusion that spin waves significantly hamper the switching dynamics of small magnetic p-bits.}

{Next, we address micromagnetic calculations of the RTN frequency $f_{s,exch-dip}$ under parametric excitation by applying a
uniform microwave field $P_{p}\cos(2\omega_{0}t)/\gamma$ or $P_{p^{\prime}}\cos(2\omega_{sw}t)/\gamma$  polarized parallel to the in-plane easy axis ($\hat{x}$), where
$\omega_{0}\approx1\,$GHz is the Kittel mode frequency. We chose a spin wave frequency $\omega_{sw}=3.5\,$GHz based on the absorption spectrum {in Fig. \ref{fig3}(b)} of a microwave field polarized normal to the plane. In order to interact with both odd and even magnon modes, we chose it to be constant as a function of \textit{y} but linearly increasing from zero at $x=-r$ to $2\,$mT at $x=r$. The peak at $\omega_{sw}\approx3.5\,$GHz is the lowest spin wave resonance above the Kittel mode $\omega_0$. The calculated $f_{s,\textrm{norm}}$
as a function of $P_{p}$ and
$P_{p^{\prime}}$ in Fig. \ref{fig3}(c), where `norm' in the subscript refers to the
normalization to $P_{p}=P_{p^{\prime}}=0$, agrees with our
predictions: $f_{s,\textrm{norm}}$ monotonically increases with $P_{p}$ when
$P_{p^{\prime}}=0$. Results for even larger $P_{p}$ are contaminated by coherent ringing and not shown. When increasing $P_{p^{\prime}}$ while $P_{p}=0$, $f_{s,\textrm{norm}}$ drops after an initial 
slight increase. At higher $P_{p^{\prime}}$, the switching events
become rare and estimates are statistically unreliable. Figure \ref{fig3}(d) shows the autocorrelation function for the
excitation powers indicated by arrows in Fig. \ref{fig3}(c). The trends of the autocorrelation decay rate agree with that of $f_{s,\textrm{norm}}$ indicating that coherent ringing is not an issue.}

The micromagnetic results in Fig. \ref{fig3}(c) also expose the limits of our simple model [see Fig. \ref{fig2}(b)]: (i) The strength
$P_{p^{\prime}}$ required to significantly modulate the RTN is much larger than $P_{\pm\mathcal{K}}$
because the driving force of the magnons in the micromagnetics is effectively $P_{p^{\prime}}\mathcal{E}_{\pm\mathcal{K}%
}c_{\mathcal{K}}c_{-\mathcal{K}}$
so $P_{p^{\prime}}$ must compensate for the relatively small ellipticity $\mathcal{E}%
_{\pm\mathcal{K}}\propto u_{\pm\mathcal{K}}v_{\pm\mathcal{K}}$ not included in the model. (ii) The initial acceleration of the RTN with increasing $P_{p^{\prime}}$ in the micromagnetics [see blue dots in Fig. \ref{fig3}(c)] hints at a direct Kittel mode excitation, which is possible because its larger ellipticity can partly compensate the spectral separation but not included in the model, thereby causing an initial increase with $P_{p^{\prime}}$. Only at larger $P_{p^{\prime}}$, the suppression of $f_{s}$ by the excited spin waves dominate. (iii) The micromagnetic calculations do not observe the increase after the downturn in the model  $f_{s,\textrm{norm}}$ in Fig. \ref{fig2}(b) for larger $P_{\pm\mathcal{K}}$, signaling another breakdown of the single pair approximation in the strongly driven magnet.

{Figure \ref{fig3}(e) shows $f_{s,\mathrm{norm}}$ as a function of the microwave frequency $2\omega_{sw}$ for a fixed $P_{p'}/(\alpha_{G}\omega_{sw})=12$. In Fig. \ref{fig3}(e), we use a finer mesh of $2\,\textrm{nm}\times 2\,\textrm{nm}$. The smaller number of runs (10) leads to larger statistical error bars. $f_{s,\mathrm{norm}}$ is substantially suppressed close to the two spin wave resonances resolved in  Fig. \ref{fig3}(b) as predicted by the model.}

\section{\label{cooling} Mode-selective cooling}
Phonons can be cooled by photons \cite{Aspelmeyer2014} with active feedback \cite{Mancini1998,Koch2011}, Brillouin scattering \cite{Tomes2011}, and radiation pressure interaction \cite{Cohadon1999,Aspelmeyer2014}. The radiation pressure interaction $ga^{\dagger}a(b+b^{\dagger})$, where $g$ is the bare interaction strength and \(a\) the photon annihilation operator, can cool a boson $b$ with frequency $\omega_{b}$ when the number of photons $n_{th,a}<n_{th,a}$ and its dissipation rate $ \xi_{a}\ll \omega_b$. An incoming photon beam that is red shifted from the resonance frequency $\omega_{a}$ by $\omega_{b}$ annihilates a \(b\) boson by anti-Stokes scattering. The radiation pressure can also cool the Kittel mode \cite{Osada2016,Zhang2016,Haigh2016,Kusminskiy2016,Sharma2017}.  Here we are interested in the  $\pm\mathcal{K}$ modes in MTJs but the radiation pressure $g a^{\dagger}_{p}a_{p}(c_{\mathcal{K}}+c_{-\mathcal{K}}^{\dagger})$, where $a_{p}$ annihilates a plane-wave photon, does not conserve linear momentum. An alternative feedback cooling with a ``beam-splitter" interaction \(g_{bs}(a_p c_\mathcal{K}^\dagger + \mathrm{H.c.})\) suffers from the same issue. Here we overcome these limitations by an in-situ cooling mechanism based on the four-magnon interaction. 

\subsection{\label{nonlin_cool_model}Model}
The easy axis anisotropy $\mathcal{H}_{ea}=\gamma H_{K_{e}}m_{x}^2$ expanded up to fourth order in the magnon amplitudes $c_{\vec{k}}$, reads
\begin{align}
\mathcal{H}_{ea}&\approx \sum_{\vec{k}}\left[A_{1}c_{\vec{k}}^{\dagger}c_{\vec{k}}+(A_{2}c_{\vec{k}}c_{-\vec{k}}+\mathrm{H.c.})+\right.\nonumber\\
&\left.c_{\vec{k}}^\dagger c_{\vec{k}}(A_{3}c_{0}+\mathrm{H.c.})+\sum_{\vec{k}^{\prime}}c_{\vec{k}}^\dagger c_{\vec{k}}(A_{4}c_{\vec{k}^{\prime}}c_{-\vec{k}^{\prime}}+\mathrm{H.c.})\right].\label{eq_cool1}
\end{align}
The constants $A_{1-4}\propto H_{K_{e}}$ can be modulated by the voltage-controlled magnetic anisotropy (VCMA) effect, see below in Sec. \ref{exp}. \(A_{2}\) term in Eq. (\ref{eq_cool1}) can parametrically excite magnon modes by an ac voltage. The term with \(A_{3}\) can cool only the uniform mode \(c_0\). Cooling the $\pm\mathcal{K}$ becomes possible by the \(A_4\) interaction which in its full form also includes contributions from the exchange interaction \cite{Krivosik2010}. By keeping only terms that are relevant for cooling we arrive at a minimum model with two magnon modes \(\vec{k}'\)and  \( \vec{k}\), where \(\vec{k}'\) is the mode \(\mathcal{K} \) to be cooled by the excitation of magnons with wave vector \( \vec{k}\):
\begin{align}
&\mathcal{H}_{cool}=\Delta c_{\vec{k}}^{\dagger}c_{\vec{k}}+\omega_{\vec{k}^{\prime}}c_{\vec{k}^{\prime}}^{\dagger}c_{\vec{k}^{\prime}}+\nonumber\\
&A_{4}c_{\vec{k}}^\dagger c_{\vec{k}}(c_{\vec{k}^{\prime}}c_{-\vec{k}^{\prime}}+H.c.)+H_{d,\vec{k}},\label{eq_cool2}
\end{align} 
where $\Delta=\omega_{\vec{k}}-\omega_{d}$, and $H_{d,\vec{k}}=P_{\textrm{res}}c_{\vec{k}}+\mathrm{H.c.}$ resonantly drives the $\vec{k}$ mode by a spatially inhomogeneous field with frequency $\omega_{d}$ and excitation amplitude $P_{\textrm{res}}$. Employing parametric excitation using the $A_{2}$ terms rather than resonant drive can achieve the same cooling, and will be discussed in Ref. \cite{Elyasi2026}. By linearizing $c_{\vec{k}}^{\dagger}c_{\vec{k}}$ around the mean field of the $\vec{k}$ mode, $c_{\vec{k}}=\alpha_{\vec{k}}+\delta c_{\vec{k}}$ , where $\alpha_{\vec{k}}=P_{\textrm{res}}/(\xi_{k}+i\Delta)$ and $\delta c_{\vec{k}}$ is the fluctuating field operator
\begin{align}
&\mathcal{H}_{cool,eff}=\Delta \delta c_{\vec{k}}^{\dagger}\delta c_{\vec{k}}+\omega_{\vec{k}^{\prime}}c_{\vec{k}^{\prime}}^{\dagger}c_{\vec{k}^{\prime}}+\nonumber\\
&(G\delta c_{\vec{k}}+G^{\ast}\delta c_{\vec{k}}^\dagger)(\mathfrak{c}_{\vec{k}^{\prime}}+\mathrm{H.c.}),\label{eq_cool3}
\end{align} 
where $\mathfrak{c}_{\vec{k}^{\prime}}=c_{\vec{k}^{\prime}}c_{-\vec{k}^{\prime}}$, $G=A_{4}\alpha_{\vec{k}}$ and we chose the phase $\mathrm{Im} \, \alpha_{\vec{k}}=\mathrm{Im} \, G =0$. 

\begin{figure}[ptb]
\includegraphics[width=0.5\textwidth]{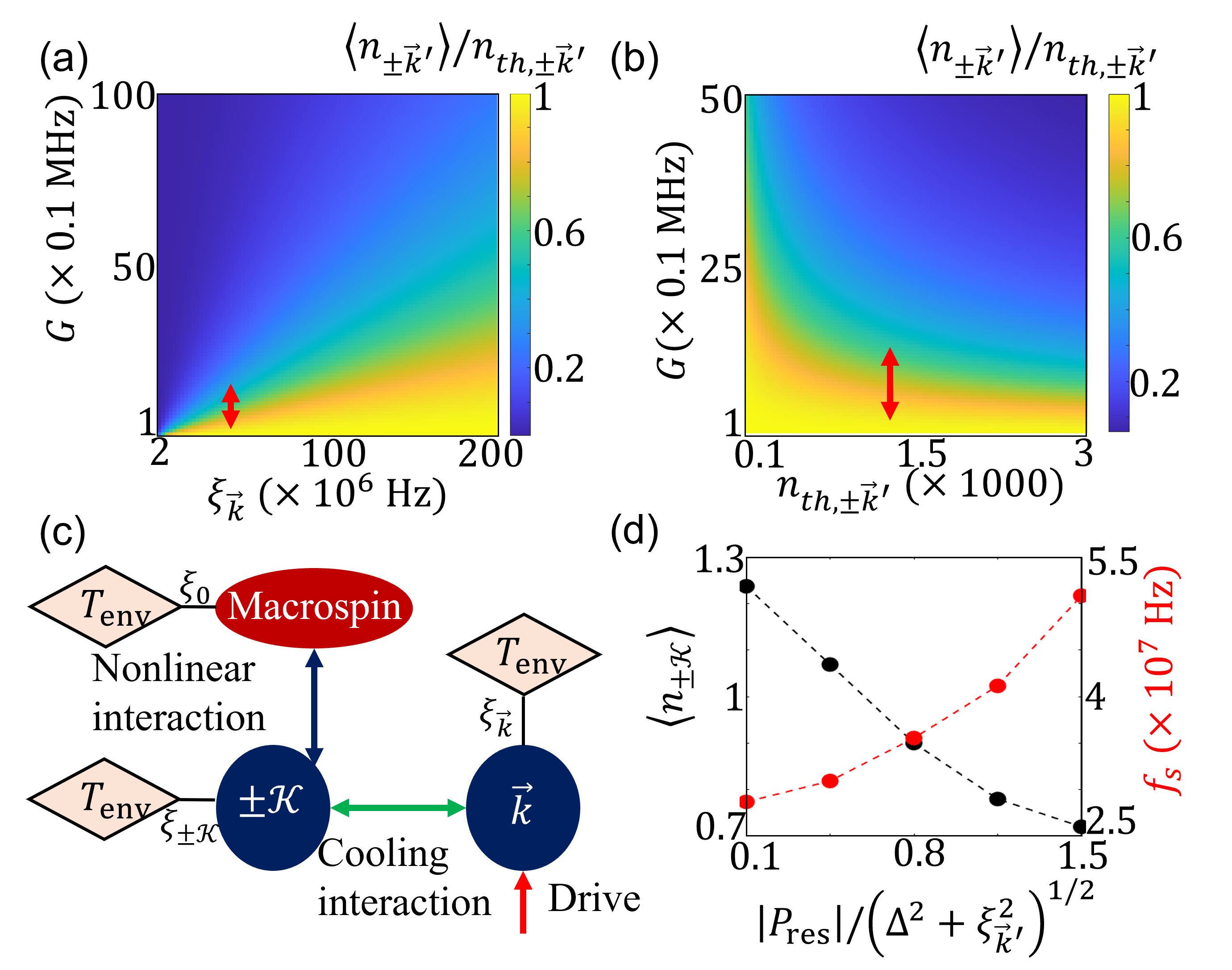}
\caption{Magnon cooling and magnetization RTN. (a) {$\langle n_{\pm\vec{k}^{\prime}}%
\rangle/n_{th,\pm\vec{k}^{\prime}}$ ($<1$ implies cooling of the mode \(\vec{k}'=\mathcal{K}\)) as} a function of the dissipation rate $\xi_{\vec{k}}$ of the driven magnon with wave vector \(\vec{k}\) and the interaction strength $G$. (b) Cooling rate as a function of the initial thermal occupation $n_{th,\pm\vec{k}^{\prime}}$, and $G$. (c) The model for magnetization RTN in the presence of the nonlinear interaction and the four-magnon cooling. (d) $\langle n_{\pm\mathcal{K}}\rangle$ (black dots) and $f_{s}$ (red dots) as a function of the drive amplitude $P_{res}$ ($G\propto P_{res}$). The red double-sided arrows in (a) and (b) indicated the range of $G$ spanned by the drive parameter $P_{res}$ in (d).}
\label{fig4}%
\end{figure}

\subsection{\label{nonlin_cool_results}Non-linear cooling and RTN}
The EOMs for $\langle n_{\vec{k}}\rangle$, $\langle n_{\vec{k}^{\prime}}\rangle$, $\langle\delta c_{\vec{k}}^{\dagger}\mathfrak{c}_{\vec{k}^{\prime}}\rangle$, $\langle\mathfrak{c}_{\vec{k}^{\prime}}^{\dagger}\mathfrak{c}_{\vec{k}^{\prime}}\rangle$, and their complex conjugates form a closed system of equations. The Lindblad master equation for the Hamiltonian in Eq. (\ref{eq_cool3}) and linear dissipation, leads to an equation for $\langle n_{\vec{k}^{\prime}}\rangle$
\begin{align}
&Q_{1}\langle n_{\vec{k}^{\prime}}\rangle^2+Q_{2}\langle n_{\vec{k}^{\prime}}\rangle+Q_{3}=0,\label{eq_cool4}
\end{align}
where
\begin{align}
&Q_{1}=-\frac{B}{2}-\frac{\xi_{\vec{k}^{\prime}}}{\xi_{\vec{k}}}B,\\
&Q_{2}=Bn_{th,\vec{k}}+\frac{\xi_{\vec{k}^{\prime}}}{\xi_{\vec{k}}}Bn_{th,\vec{k}^{\prime}}-\frac{\xi_{\vec{k}^{\prime}}}{G},\\
&Q_{3}=\frac{\xi_{\vec{k}^{\prime}}}{G}n_{th,\vec{k}^{\prime}},\\
&B=\frac{2G(2\xi_{\vec{k}^{\prime}}+\xi_{\vec{k}})}{(\Delta-2\omega_{\vec{k}^{\prime}})^2+(2\xi_{\vec{k}^{\prime}}+\xi_{\vec{k}})^2}.\label{eq_cool5}
\end{align}
At the anti-Stokes scattering resonance $\Delta=2\omega_{\vec{k}^{\prime}}$ and under the condition $\{G,\xi_{\vec{k}^{\prime}}\}\ll\xi_{\vec{k}}$, i.e., relatively weak drive, the solution of Eq. (\ref{eq_cool4}) simplifies to
\begin{align}
\langle n_{\vec{k}^{\prime}}\rangle=n_{th,\vec{k}^{\prime}}-\frac{G^2}{\xi_{\vec{k}}\xi_{\vec{k}^{\prime}}}n_{th,\vec{k}^{\prime}}(n_{th,\vec{k}^{\prime}}-2n_{th,\vec{k}}).\label{eq_cool9}
\end{align}
When \(n_{th,\vec{k}}\) is small, $\langle n_{\vec{k}^{\prime}}\rangle$ is reduced from its equilibrium value $n_{th,\vec{k}^{\prime}}$,  which is equivalent to cooling. In contrast to the radiation pressure mechanism, the cooling rate depends on $n_{th,\vec{k}^{\prime}}$ or $T_{\textrm{env}}$, and is therefore (for fixed $G$) more efficient at higher temperatures. Figures \ref{fig4}(a) and (b) illustrate the parameter dependence of the cooling based on the numerical solution of Eq. (\ref{eq_cool4}).

With \(\vec{k}^{\prime}\rightarrow \pm \mathcal{K}\), we still have to specify the mode with wave vector \(\vec{k}\).  Cooling is efficient when its thermal occupation $n_{th,\vec{k}}\ll n_{th,\pm\mathcal{K}}$, hence $\omega_{\vec{k}}\gg\omega_{\pm\mathcal{K}}$ and  $\xi_{\vec{k}}/\xi_{\pm\mathcal{K}}\approx\omega_{\vec{k}}/\omega_{\pm\mathcal{K}}\gg 1$ for constant Gilbert damping.
 We discuss the cooling by parametric excitation via $A_{2}$  in Ref. \cite{Elyasi2026}.
We then include this mode ($\vec{k}$) into the RTN model of Ref. \cite{Elyasi2024} [see Fig. \ref{fig4}(c)]. Figure \ref{fig4}(d) shows that $\tau_{s}$ and $\langle n_{\pm\mathcal{K}}\rangle$ indeed decrease with increasing $\vert P_{\textrm{res}}\vert$. 
Here we adopted 
$\omega_{\vec{k}}=8\omega_{\pm\mathcal{K}}$ 
($\sim$ the frequency of the spin wave mode band
with five nodes in a \(r=20\)\,nm sample), and $A_{4}\approx2\times10^{4}$  from Eq. (64) of Ref. \cite{Krivosik2010}. A smaller $\alpha_{G}=0.005$   reduces the computational effort caused by adding the additional mode without affecting the trends in $f_{s}$.

\section{\label{exp} Experimental considerations}

Magnons have been routinely excited by photons or VCMA \cite{Zhu2012,Rana2017,Verba2014,Chen2017,Rana2019}. Nano-striplines \cite{Sheng2023} are suitable microwave sources to excite the magnetization in MTJs. The Zeeman interaction $\mathcal{H}_{ex,ph}=-\gamma\vec{h}_{mw}\cdot\vec{m}$  for a spatially uniform field \(\vec{h}_{mw}=(h_x,h_y,h_z)\) and an easy axis simplifies to
\begin{align}
&\mathcal{H}_{ex,ph}=P_{res,ph}c_{0}+\sum_{\vec{k}}(P_{p,ph,\vec{k}}c_{\vec{k}}c_{-\vec{k}})+H.c.,\nonumber\\
&P_{res,ph}=\gamma[\frac{h_{x}}{2}((u_{0}+v_{0})\sin\phi_{s}-i(u_{0}-v_{0})\cos\theta_{s}\cos\phi_{s})+\nonumber\\
&\frac{h_{y}}{2}((u_{0}+v_{0})\cos\phi_{s}-i(u_{0}-v_{0})\cos\theta_{s}\sin\phi_{s})+\nonumber\\
&\frac{h_{z}}{2}(i(u_{0}-v_{0})\sin\theta_{s})],\nonumber\\
&P_{p,ph,\vec{k}}=2\gamma u_{\vec{k}}v_{\vec{k}}[h_{x}\sin\theta_{s}\cos\phi_{s}\nonumber\\
&+h_y\sin\theta_{s}\sin\phi_{s}+h_z\cos\theta_{s}],\label{eq_ex1}
\end{align}
where $\theta_{s}$ and $\phi_{s}$ are polar and azimuthal angles of $\vec{m}_{eqb}$ with respect to the $\hat{z}$-axis, while $u_{\vec{k}}$ and $v_{\vec{k}}$ parameterize the ellipticity of the magnon modes after a Bogoliubov transformation to magnon annihilation and creation operators $a_{\vec{k}}=u_{\vec{k}}c_{\vec{k}}+v_{\vec{k}}c_{-\vec{k}}^{\dagger}$ and $a^{\dagger}_{-\vec{k}}=v^{\ast}_{\vec{k}}c_{\vec{k}}+u_{\vec{k}}c_{-\vec{k}}^{\dagger}$, respectively. Parametric excitation is not possible when precession is circular, or $v_{\vec{k}}=0$ and $u_{\vec{k}}=1$. The ellipticity of the Kittel mode does not include the exchange interaction $\propto\vert \vec{k}\vert ^2$ that suppresses $u_{\vec{k}}v_{\vec{k}}$ and therefore parametric excitation of higher frequency modes. Similar arguments hold for the parametric excitation of MTJs by ac electric fields \cite{Chen2017}. When the easy axis is fixed along $\hat{z}$, modulating the hard axis anisotropy $K_{h}$ along $\hat{x}$ and maintaining $\vec{m}_{eqb}\perp\hat{x}$ at an angle $\theta_{s}$ with the easy axis $\|\hat{z}$, 
\begin{align}
&\mathcal{H}_{ex,an,h}=\gamma H_{K_{h}}(\vec{m}\cdot\hat{x})^2=\sum_{\vec{k}}(P_{p,an,h,\vec{k}}c_{\vec{k}}c_{-\vec{k}})+H.c.,\nonumber\\
&P_{p,an,h,\vec{k}}=\gamma H_{K_{h}}u_{\vec{k}}v_{\vec{k}},\label{eq_ex2}
\end{align}
On the other hand, modulating the easy axis anisotropy in the same configuration,
\begin{align}
&\mathcal{H}_{ex,an,e}=\gamma H_{K_{e}}(\vec{m}\cdot\hat{z})^2=P_{res,an,e}c_{0}+\nonumber\\
&\sum_{\vec{k}}(P_{p,an,e,\vec{k}}c_{\vec{k}}c_{-\vec{k}})+H.c.,\nonumber\\
&P_{res,an,e}=2\gamma H_{K_{e}}[-i\sqrt{2}\sin\theta_{s}\cos\theta_{s}(u_{0}-v_{0})],\nonumber\\
&P_{p,an,e,\vec{k}}=\gamma H_{K_{e}} u_{\vec{k}}v_{\vec{k}}[-4\cos^2\theta_{s}-\sin^2\theta_{s}].\label{eq_ex3}
\end{align}
In experiments, the ac voltage over a tunnel barrier affects the OOP anisotropy so for an in-plane equilibrium magnetization $\mathcal{H}_{ex,an,h}$ is appropriate. For perpendicularly magnetized samples, $\mathcal{H}_{ex,an,e}$ applies, and $P_{p,an,e,\vec{k}}\neq 0$ in the presence of in-plane crystalline or shape anisotropies.

The cooling mechanism in Sec. \ref{cooling} relies on the excitation of a magnon mode $\vec{k}$ at preferably few tens of GHz higher frequencies than that of the $\pm\mathcal{K}$ magnons. The decreased ellipticity can be compensated by higher excitation power, less dissipation, and larger anisotropies [see Fig. \ref{fig4}(a)].

\section{Conclusion}
We propose and formulate active methods to control magnetization RTN in MTJs. We analytically predict and numerically validate RTN attempt times that increase with the parametric excitation power of the macrospin. However, the drive also reduces the energy barrier, which wins against the longer attempt times, accelerating the switching rate. An experimental verification of the predicted reduced RTN frequencies under parametric spin-wave excitations would be helpful to assess the role of non-linearities in p-bit devices. We also propose an in-situ cooling by high-frequency magnons that could increase the RTN frequency by an order of magnitude with the current technology, using parametric excitation of MTJs by microwaves or ac gate voltages.  Future work should address complications such as magnetization reversal assisted by domain walls which should become increasingly important for MTJs with larger diameter \cite{Braun1993,Krause2009,Soumah2024}. 

\acknowledgments We acknowledge support by JSPS KAKENHI (Grants No. 21K13847, No. 22H04965, No. 20H02178, No. 23KK0092, No. 24H02231, No. 24H02235, and No. 24H00039), JST PRESTO (Grant No. JPMJPR21B2), MEXT X-NICS (Grant No. JPJ011438), and Cooperative Research Projects of RIEC. This work was supported by Japan Science and Technology Agency (JST) as part of Adopting Sustainable Partnerships for Innovative Research Ecosystem (ASPIRE), Grant Number JPMJAP26A1.

\end{document}